\documentclass[twocolumn]{aa}
\usepackage{graphicx}
\usepackage{txfonts}
\usepackage{natbib}
\bibpunct{(}{)}{;}{a}{}{,}

\begin{document}

\title{
The HARPS search for southern extra-solar planets\thanks{Based on observations made with the HARPS instrument on the ESO 3.6 m telescope at La Silla Observatory under programme ID 072.C-0488(E).}
}

\subtitle{III. Three Saturn-mass planets around HD 93083, HD 101930 and HD 102117}

\author{
C.~Lovis\inst{1}
\and M.~Mayor\inst{1}
\and F.~Bouchy\inst{2}
\and F.~Pepe\inst{1}
\and D.~Queloz\inst{1}
\and N.C.~Santos\inst{3}\fnmsep\inst{1}
\and S.~Udry\inst{1}
\and W.~Benz\inst{4}
\and J.-L.~Bertaux\inst{5}
\and C.~Mordasini\inst{4}
\and J.-P.~Sivan\inst{2}
}

\offprints{C. Lovis}

\institute{
Observatoire de Gen\`eve, 51 ch. des Maillettes, 1290 Sauverny, Switzerland\\
\email{christophe.lovis@obs.unige.ch}
\and
Laboratoire d'Astrophysique de Marseille, Traverse du Siphon, 13013 Marseille, France
\and
Centro de Astronomia e Astrof\'isica da Universidade de Lisboa, Observat\'orio Astron\'omico de Lisboa, Tapada da Ajuda, 1349-018 Lisboa, Portugal
\and
Physikalisches Institut Universit\"at Bern, Sidlerstrasse 5, 3012 Bern, Switzerland
\and
Service d'A\'eronomie du CNRS, BP 3, 91371 Verri\`eres-le-Buisson, France
}

\date{Received 11 February 2005 / Accepted 16 March 2005}

\abstract{
We report on the detection of three Saturn-mass planets discovered with the HARPS instrument. HD~93083 shows radial-velocity (RV) variations best explained by the presence of a companion of 0.37 $M_{\mathrm{Jup}}$ orbiting in 143.6 days. HD~101930~b has an orbital period of 70.5 days and a minimum mass of 0.30 $M_{\mathrm{Jup}}$. For HD~102117, we present the independent detection of a companion with $m_2 \sin{i}$ = 0.14 $M_{\mathrm{Jup}}$ and orbital period $P$ = 20.7 days. This planet was recently detected by \citet{tinney04}. Activity and bisector indicators exclude any significant RV perturbations of stellar origin, reinforcing the planetary interpretation of the RV variations. The radial-velocity residuals around the Keplerian fits are 2.0, 1.8 and 0.9 m~s$^{-1}$ respectively, showing the unprecedented RV accuracy achieved with HARPS. A sample of stable stars observed with HARPS is also presented to illustrate the long-term precision of the instrument. All three stars are metal-rich, confirming the now well-established relation between planet occurrence and metallicity. The new planets are all in the Saturn-mass range, orbiting at moderate distance from their parent star, thereby occupying an area of the parameter space which seems difficult to populate according to planet formation theories. A systematic exploration of these regions will provide new constraints on formation scenarios in the near future.
}

\maketitle

\section{Introduction}

The HARPS instrument is the new ESO high-resolution ($R$ = 115,000) fiber-fed echelle spectrograph, mainly dedicated to planet search programmes and asteroseismology. It has already proved to be the most precise spectro-velocimeter to date, reaching an instrumental RV accuracy of $\sim$1 m~s$^{-1}$ \citep{mayor03,pepe04,santos04a}, and even better on a short-term basis. This opens a new field in the search for extrasolar planets, allowing the detection of companions of a few Earth masses around solar-type stars. Indeed, the increase of the planet frequency towards very low masses is confirmed by the recent discovery of Neptune-mass exoplanets \citep{mcarthur04,santos04a,butler04}. Moreover, the combination of CORALIE \citep{queloz00} and HARPS data, extending over several years, will allow the characterization of long-period planets and complex planetary systems. The low-mass and long-period regions of the exoplanet parameter space will therefore be under close scrutiny in the coming months and years. This will improve our knowledge of the planet distribution in the mass-period diagram and will allow comparisons with theoretical predictions \citep[see for example][]{armitage02,ida04,alibert05}.

In this paper we present the discovery of three planetary companions to the stars HD~93083, HD~101930 and HD~102117. Interesting characteristics of these planets include a mass in the Saturn-mass regime and below, and an orbital distance to the star of 0.1-0.5 AU. According to recent planet formation scenarios \citep{ida04}, they are situated in a region of the mass-period diagram that seems difficult to populate. The systematic exploration of this mass and distance regime is therefore an important test for the core-accretion and migration theories of planet formation.

This paper is structured as follows. Section 2 describes the observations, the data reduction process and discusses the long-term precision of the HARPS measurements. The physical properties of the parent stars are presented in Sect. 3, followed by the RV measurements and orbital solutions in Sect. 4. We discuss in the last section the characteristics of these new planets in the context of the already-known properties of exoplanets and highlight some theoretical questions that might be answered by radial-velocity surveys in the near future.

\section{Observations and measurement precision}

HD~93083, HD~101930 and HD~102117 are all members of the HARPS high-precision RV sample. The aim of this survey is to obtain RV measurements with photon errors below 1 m~s$^{-1}$ to detect very low-mass extrasolar planets. The stars in this sample have been selected from the CORALIE planet search database for being non-evolved, having low projected rotational velocity ($v \sin{i} <$ 3 km~s$^{-1}$) and exhibiting low activity levels ($\log R'_{\mathrm{HK}} <$ -4.7). These criteria should eliminate most stars showing large intrinsic RV variations and select only very quiet, chromospherically inactive stars, so that stellar RV jitter does not hide possible planetary signals down to 1-2 m~s$^{-1}$. Obviously, the photon noise error on the radial velocity must also remain below 1 m~s$^{-1}$. For each HARPS spectrum we compute the photon-limited, ultimate RV precision using the formulae given by \citet{connes85} and \citet{bouchy01}. HARPS typically delivers 0.5-1 m~s$^{-1}$ photon-limited accuracy at a S/N ratio of $\sim$100 per pixel (0.8 km~s$^{-1}$) at 550 nm on solar-type stars. The exact number mainly depends on the depth of absorption lines and the projected rotational velocity.

The radial velocities for HD~93083, HD~101930 and HD~102117 have been obtained with the standard HARPS reduction pipeline, based on the cross-correlation with a stellar template, the precise nightly wavelength calibration with ThAr spectra and the tracking of instrumental drifts with the simultaneous ThAr technique \citep{baranne96}. Of particular interest is the fact that the nightly instrumental drifts always remain below 1 m~s$^{-1}$ for HARPS due to the high instrumental stability.

\begin{figure}
\centering
\includegraphics[angle=-90,width=9cm]{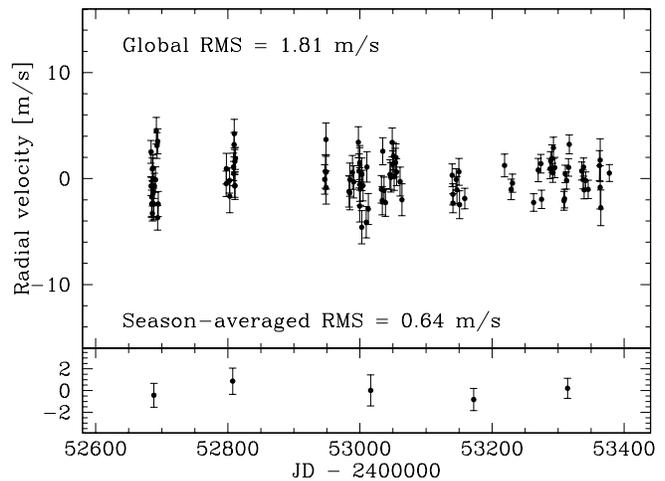}
	\caption{Radial velocities for a sample of 12 stable stars from the HARPS high-precision programme, observed over about 700 days. For each star, the mean RV has been subtracted. The overall RMS of 1.81 m~s$^{-1}$ is made of different contributions, among which photon noise ($\sim$0.8 m~s$^{-1}$), calibration noise ($\sim$0.8 m~s$^{-1}$), guiding errors (0.2-0.6 m~s$^{-1}$) and stellar acoustic modes (0.5-2.0 m~s$^{-1}$). Seasonal averages have been computed (bottom part). The absence of long-term drifts and the dispersion of only 0.64 m~s$^{-1}$ illustrate the long-term stability of ThAr reference lamps.}
	\label{FigLongTerm}
\end{figure}

To assess the long-term precision of the instrument, we selected a sample of 12 stable stars, which have been followed regularly over the past two years. These stars include HD~55 (K5V, $V$=8.49), HD~1581 (F9V, $V$=4.23), HD~6673 (K1V, $V$=8.84), HD~20794 (G8V, $V$=4.26), HD~28471 (G5V, $V$=7.89), HD~30278 (G8V, $V$=7.61), HD~44594 (G4V, $V$=6.61), HD~82342 (K3/K4V, $V$=8.31), HD~114853 (G2V, $V$=6.93), HD~162396 (F8V, $V$=6.19), HD~196761 (G8/K0V, $V$=6.36) and HD~215152 (K0V, $V$=8.11). For each of these stars, several measurements have been gathered, spanning at least 300 days, and the mean RV was subtracted to enable a comparison between different stars. The first data points were acquired during the commissioning periods of the instrument in February and June 2003. Between then and the more recent measurements, many modifications have been made to the instrument, especially the exchange of the ThAr reference lamps. In spite of that, the global RV dispersion of the measurements amounts to 1.81 m~s$^{-1}$ (see Fig.\ref{FigLongTerm}). This includes many different effects, which can be estimated as follows.

\begin{itemize}

\item The mean photon error, computed according to \citet{bouchy01}, is 0.8 m~s$^{-1}$.

\item The zero-point wavelength calibration, given by nightly ThAr reference spectra, is presently determined with an accuracy of $\sim$0.8 m~s$^{-1}$.

\item Guiding errors are responsible for 0.2-0.6 m~s$^{-1}$ of extra dispersion, due to the non-optimal tuning of the guiding software. This issue has now been solved and the guiding noise should remain below 0.3 m~s$^{-1}$ in the future.

\item The largest contribution to the global RV dispersion comes from the star itself, mainly due to acoustic modes (p-modes). For our 12 stars we estimate this contribution to be 0.5-2.0 m~s$^{-1}$, depending on spectral type and evolutionary status \citep[see][]{mayor03,bouchy05}. The observational strategy has been recently optimized to minimize these stellar oscillations, essentially by integrating over long enough periods to cover more than 1-2 oscillations (3-15 minutes, depending on spectral type).

\end{itemize}

Adding all these effects quadratically, we obtain a total dispersion that is very close to the measured value of 1.81 m~s$^{-1}$ (the exact number depending on the adopted mean value for the oscillation noise). This leaves only little space (probably less than 1 m~s$^{-1}$) for other star-related variations, such as the presence of undetected planetary companions, star spots and activity-induced jitter.

Among the above-mentioned error sources, three of them should be of less importance in the future, permitting an even better precision. Firstly, guiding errors are no longer a limiting factor. Secondly, the new observing strategy should allow us to maintain the stellar oscillation "noise" below 0.5 m~s$^{-1}$ for K-dwarfs and 1.5 m~s$^{-1}$ for slightly evolved, early G-dwarfs (worst case). Indeed, comparisons between measurements on a given star taken before and after the strategy change have already shown a significant reduction of the RV dispersion. Finally, improvements in the wavelength calibration are still ongoing, in particular the elaboration of a more accurate list of ThAr laboratory wavelengths, which will lead to a lower calibration noise. The improved wavelength solutions will then be applied to all data (past and future), ensuring an even better night-to-night stability.

To study the behavior of reference ThAr spectra over time, we have computed seasonal averages (see Fig.\ref{FigLongTerm}, bottom). The dispersion of these seasonal mean values amounts to 0.64 m~s$^{-1}$, showing that ThAr lamps are able to deliver an absolute wavelength reference at that level of precision over long periods of time, even in the case of lamp exchange or lamp current variations (both occurred during the first year of HARPS operation). In conclusion, we are therefore very confident of the HARPS ability to maintain a long-term accuracy at the level of 1 m~s$^{-1}$.

\section{Parent star characteristics}

The basic properties of HD~93083 (K3V, $V$=8.30), HD~101930 (K1V, $V$=8.21) and HD~102117 (G6V, $V$=7.47) are given by the Hipparcos catalogue \citep{esa97}. Physical properties have been derived following the method by \citet{santos01,santos04b}, using a standard local thermodynamical equilibrium (LTE) analysis. This study provides precise values for the effective temperatures, metallicities and surface gravity.

\begin{table}
\caption{Observed and inferred stellar parameters for the planet-hosting stars presented in this paper.}
\label{TableStars}
\centering
\begin{tabular}{l l c c c}
\hline\hline
\multicolumn{2}{l}{\bf Parameter} & \bf HD 93083 & \bf HD 101930 & \bf HD 102117 \\
\hline
Sp & & K3V & K1V & G6V \\
$V$ & [mag] & 8.30 & 8.21 & 7.47 \\
$B-V$ & [mag] & 0.945 & 0.908 & 0.721 \\
$\pi$ & [mas] & 34.60 & 32.79 & 23.81 \\
$M_V$ & [mag] & 6.00 & 5.79 & 4.35 \\
$T_{\mathrm{eff}}$ & [K] & 4995 $\pm$ 50 & 5079 $\pm$ 62 & 5672 $\pm$ 22 \\
log $g$ & [cgs] & 4.26 $\pm$ 0.19 & 4.24 $\pm$ 0.16 & 4.27 $\pm$ 0.07 \\
$\mathrm{[Fe/H]}$ & [dex] & 0.15 $\pm$ 0.06 & 0.17 $\pm$ 0.06 & 0.30 $\pm$ 0.03 \\
$L$ & [$L_{\odot}$] & 0.41 & 0.49 & 1.57 \\
$M_*$ & [$M_{\odot}$] & 0.70 $\pm$ 0.04 & 0.74 $\pm$ 0.05 & 1.03 $\pm$ 0.05 \\
$v\sin{i}$ & [km s$^{-1}$] & 0.9 & 0.7 & 1.5 \\
$\log R'_{\mathrm{HK}}$ & & -5.02 $\pm$ 0.02 & -4.99 $\pm$ 0.02 & -5.03 $\pm$ 0.02 \\
$P_{\mathrm{rot}}$ & [days] & 48 & 46 & 34 \\
\hline
\end{tabular}
\end{table}

From the colour index, the measured effective temperature and the corresponding bolometric correction, we estimate the star luminosities and we then interpolate the masses and ages in the grid of Geneva stellar evolutionary models with appropriate metal abundances \citep{schaller92,schaerer93}. The projected rotational velocity $v \sin{i}$ is also computed using the calibration of the CORALIE cross-correlation function given by \citet{santos02}. Table \ref{TableStars} gathers the photometric, astrometric, spectroscopic information and inferred quantities for HD~93083, HD~101930 and HD~102117.

The results show that the masses of these stars range from 0.7 to 1.03 $M_{\odot}$. Interestingly, they all have a metallicity higher than solar ([Fe/H] = 0.15, 0.17 and 0.30 respectively), confirming the now well-established relation between planet occurrence and metallicity \citep{santos04b}.

\begin{figure}
\centering
\includegraphics[width=9cm]{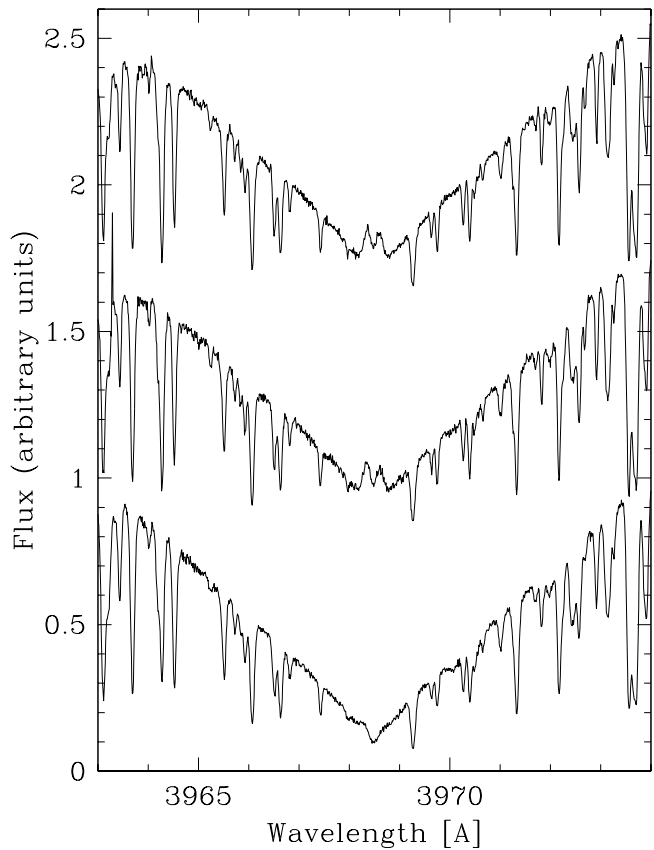}
	\caption{CaII H ($\lambda$ = 3968.47 \AA) absorption line region for HD~93083 (top), HD~101930 (middle) and HD~102117 (bottom). Reemission peaks are absent or very weak in these three stars, indicating low chromospheric activity.}
	\label{FigCaII}
\end{figure}

We also compute from the spectra an activity indicator $S_{\mathrm{HARPS}}$ measuring the re-emission flux in the CaII H and K lines. This indicator has been calibrated on the Mount Wilson scale \citep{baliunas95} using stars with well-known activity levels. We then correct for the photospheric contribution following \citet{noyes84} and obtain the well-known activity index $\log R'_{\mathrm{HK}}$. This index represents a useful tool to estimate the stellar RV jitter expected for each star due to rotational modulation of star spots or other active regions on the stellar surface \citep{saar97}. From this indicator we derive an estimation of the stellar rotation period following the calibrations by \citet{noyes84}. Although these relations have been shown not to be reliable for ages greater than $\sim$2 Gyr \citep{pace04}, they are still helpful in distinguishing between active, young stars and chromospherically quiet, old stars.

All three stars examined in this paper exhibit low chromospheric activity ($\log R'_{\mathrm{HK}}$ = -5.0, see Fig.\ref{FigCaII}). This is further supported by the non-detection of X-ray emission from these stars, despite their relatively small distance. Together with the measured small $v \sin{i}$ ($<$ 1.5 km~s$^{-1}$), this indicates very low activity-induced RV jitter ($<$ 1-2 m~s$^{-1}$). Moreover, the analysis of the bisector shape of the cross-correlation function \citep[see][]{queloz01} shows no variations in the CCF profile down to the photon noise level, giving strong support to the planetary interpretation of the RV variations.

\section{Radial-velocity data and orbital solutions}

\begin{table*}
\caption{Orbital and physical parameters for the planets presented in this paper.}
\label{TablePlanets}
\centering
\begin{tabular}{l l c c c}
\hline\hline
\multicolumn{2}{l}{\bf Parameter} & \bf HD 93083 b & \bf HD 101930 b & \bf HD 102117 b \\
\hline
$P$ & [days] & 143.58 $\pm$ 0.60 & 70.46 $\pm$ 0.18 & 20.67 $\pm$ 0.04 \\
$T$ & [JD-2400000] & 53181.7 $\pm$ 3.0 & 53145.0 $\pm$ 2.0 & 53100.1 $\pm$ 0.1 \\
$e$ & & 0.14 $\pm$ 0.03 & 0.11 $\pm$ 0.02 & 0.00 (+0.07) \\
$V$ & [km s$^{-1}$] & 43.6418 $\pm$ 0.0004 & 18.3629 $\pm$ 0.0003 & 49.5834 $\pm$ 0.0003 \\
$\omega$ & [deg] & 333.5 $\pm$ 7.9 & 251 $\pm$ 11 & 162.8 $\pm$ 3.0 \\
$K$ & [m s$^{-1}$] & 18.3 $\pm$ 0.5 & 18.1 $\pm$ 0.4 & 10.2 $\pm$ 0.4 \\
$a_1 \sin{i}$ & [10$^{-3}$ AU] & 0.239 & 0.116 & 0.019 \\
$f(m)$ & [10$^{-9} M_{\odot}$] & 0.088 & 0.042 & 0.0023 \\
$m_2 \sin{i}$ & [$M_{\mathrm{Jup}}$] & 0.37 & 0.30 & 0.14 \\
$a$ & [AU] & 0.477 & 0.302 & 0.149 \\
\hline
$N_{\mathrm{meas}}$ & & 16 & 16 & 13 \\
$Span$ & [days] & 383 & 362 & 383 \\
$\sigma$ (O-C) & [m s$^{-1}$] & 2.0 & 1.8 & 0.9 \\
\hline
\end{tabular}
\end{table*}

\subsection{HD 93083}

\begin{figure}
\centering
\includegraphics[width=9cm]{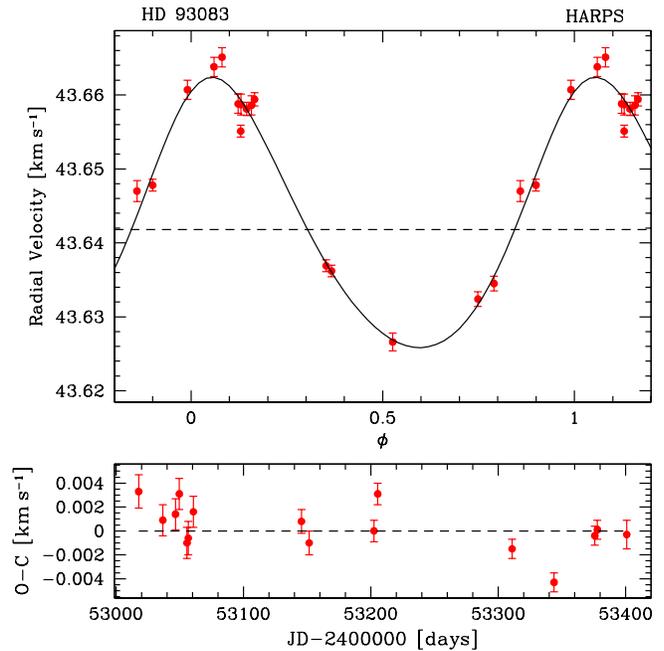}
	\caption{Phased radial velocities for HD~93083. The best Keplerian fit to the data gives a minimum mass of 0.37 $M_{\mathrm{Jup}}$ and an orbital period of 143.6 days for the planet.}
	\label{FigHD93083}
\end{figure}

Sixteen radial-velocity measurements have been obtained for HD~93083 spanning about one year. Exposure time was 15 min on average, yielding a typical S/N ratio of 130 per pixel at 550 nm. The mean photon noise error on a single data point is 0.4~m~s$^{-1}$, to which we quadratically added the corresponding calibration error (see Sect. 2). The list of all RV measurements for HD~93083 can be found in Table \ref{TableRV_HD93083}. Fig.\ref{FigHD93083} shows the radial velocities folded to a period $P$ = 143.58 days, given by the best Keplerian fit to the data. The RV semi-amplitude is $k$ = 18.3 m~s$^{-1}$ and the eccentricity $e$ = 0.14. These parameters lead to a minimum mass $m_2 \sin{i}$ = 0.37 $M_{\mathrm{Jup}}$ for the planet. All relevant data for this planet are listed in Table \ref{TablePlanets}.

The weighted rms of the residuals around the fit is 2.0 m~s$^{-1}$, which is larger than the internal errors. More data points are needed to establish whether this extra dispersion is intrinsic to the star or whether it could be explained by the presence of a third body in the system. The latter hypothesis is reinforced by the slow RV drift that seems to be present in the residuals of the radial velocities around the orbit, as can be seen in Fig.\ref{FigHD93083} (bottom).

\begin{table}
\caption{Radial-velocity values and error bars for HD~93083. All data are relative to the solar system barycenter.}
\label{TableRV_HD93083}
\centering
\begin{tabular}{c c c}
\hline\hline
\bf JD-2400000 & \bf RV & \bf Uncertainty \\
 & \bf [km~s$^{-1}$] & \bf [km~s$^{-1}$] \\
\hline
53017.84496 & 43.6470 & 0.0014 \\
53036.78231 & 43.6607 & 0.0013 \\
53046.66898 & 43.6638 & 0.0013 \\
53049.71044 & 43.6651 & 0.0013 \\
53055.74405 & 43.6588 & 0.0013 \\
53056.75374 & 43.6587 & 0.0014 \\
53060.75307 & 43.6586 & 0.0013 \\
53145.55149 & 43.6324 & 0.0010 \\
53151.57396 & 43.6345 & 0.0010 \\
53202.47605 & 43.6581 & 0.0009 \\
53205.45320 & 43.6594 & 0.0009 \\
53310.85564 & 43.6478 & 0.0009 \\
53343.85438 & 43.6551 & 0.0008 \\
53375.83553 & 43.6369 & 0.0008 \\
53377.84184 & 43.6362 & 0.0008 \\
53400.79648 & 43.6266 & 0.0012 \\
\hline
\end{tabular}
\end{table}

\subsection{HD 101930}

We have obtained 16 HARPS radial-velocity measurements of HD~101930 over a time span of $\sim$1 year. The exposure time was 15 min, yielding a typical S/N ratio of 110 per pixel at 550 nm. The mean photon noise error on a single data point is 0.5 m~s$^{-1}$, to which we quadratically added the calibration error (see Sect. 2). Table \ref{TableRV_HD101930} gives the list of all RV measurements obtained for HD~101930.

\begin{figure}
\centering
\includegraphics[width=9cm]{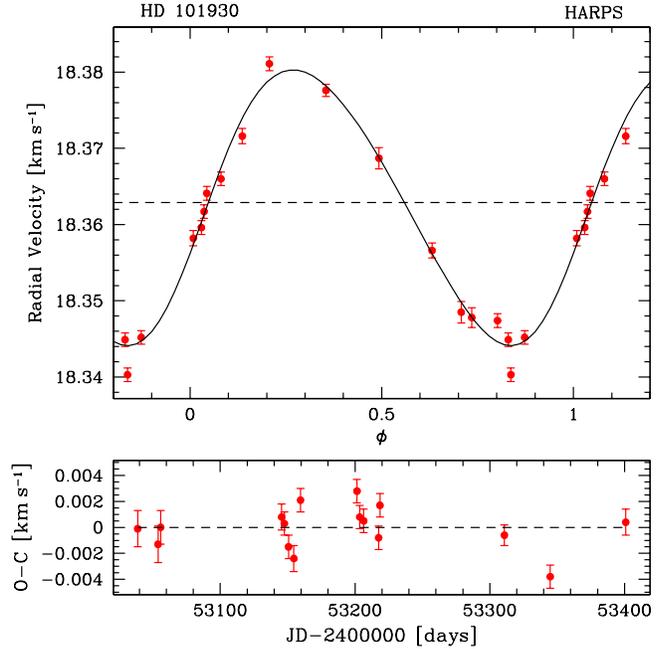}
	\caption{Phased radial velocities for HD~101930. The planet has a minimum mass of 0.30 $M_{\mathrm{Jup}}$ and an orbital period of 70.5 days.}
	\label{FigHD101930}
\end{figure}

The RV variations are best explained by a Keplerian orbit with period $P$ = 70.46 days, semi-amplitude $k$ = 18.1 m~s$^{-1}$ and eccentricity $e$ = 0.11. From these orbital parameters and the primary mass given in Table \ref{TableStars} (0.74 $M_{\odot}$), we derive a minimum mass $m_2 \sin{i}$ = 0.30 $M_{\mathrm{Jup}}$ for the companion. The residuals around the fit have a weighted rms of 1.8 m~s$^{-1}$, slightly larger than internal errors. The extra dispersion might be caused by stellar RV jitter. Future measurements will allow us to investigate this point more precisely. Fig.\ref{FigHD101930} shows the phased radial velocities for HD~101930 together with the residuals around the orbit, and Table \ref{TablePlanets} contains all relevant parameters for HD~101930~b.

\begin{table}
\caption{Radial-velocity values and error bars for HD~101930. All data are relative to the solar system barycenter.}
\label{TableRV_HD101930}
\centering
\begin{tabular}{c c c}
\hline\hline
\bf JD-2400000 & \bf RV & \bf Uncertainty \\
 & \bf [km~s$^{-1}$] & \bf [km~s$^{-1}$] \\
\hline
53038.78552 & 18.3687 & 0.0014 \\
53053.89075 & 18.3485 & 0.0014 \\
53055.83816 & 18.3478 & 0.0013 \\
53145.58216 & 18.3582 & 0.0010 \\
53147.55920 & 18.3617 & 0.0009 \\
53150.66830 & 18.3660 & 0.0009 \\
53154.58438 & 18.3716 & 0.0010 \\
53159.54707 & 18.3811 & 0.0009 \\
53201.47066 & 18.3474 & 0.0009 \\
53203.47312 & 18.3449 & 0.0009 \\
53206.46791 & 18.3452 & 0.0009 \\
53217.51591 & 18.3596 & 0.0009 \\
53218.51090 & 18.3641 & 0.0009 \\
53310.86794 & 18.3776 & 0.0009 \\
53344.86147 & 18.3403 & 0.0009 \\
53400.81929 & 18.3566 & 0.0010 \\
\hline
\end{tabular}
\end{table}

\subsection{HD 102117}

For this star we gathered 13 radial-velocity measurements with HARPS spanning about 380 days between January 2004 and January 2005. These are listed in Table \ref{TableRV_HD102117}. The spectra, acquired in 10-15 min exposures, have a mean S/N ratio of 120 per pixel, corresponding to a photon error of 0.5 m~s$^{-1}$ on this star. Note that the photon noise at a given S/N ratio is the same as for HD~101930, although HD~102117 has an earlier spectral type (G6 vs. K1), and therefore less numerous and broader spectral lines. This is probably due to the higher metallicity of HD~102117, leading to deeper absorption lines. After quadratically adding the calibration error, we obtain a total error of $\sim$0.9 m~s$^{-1}$ for each data point.

\begin{figure}
\centering
\includegraphics[width=9cm]{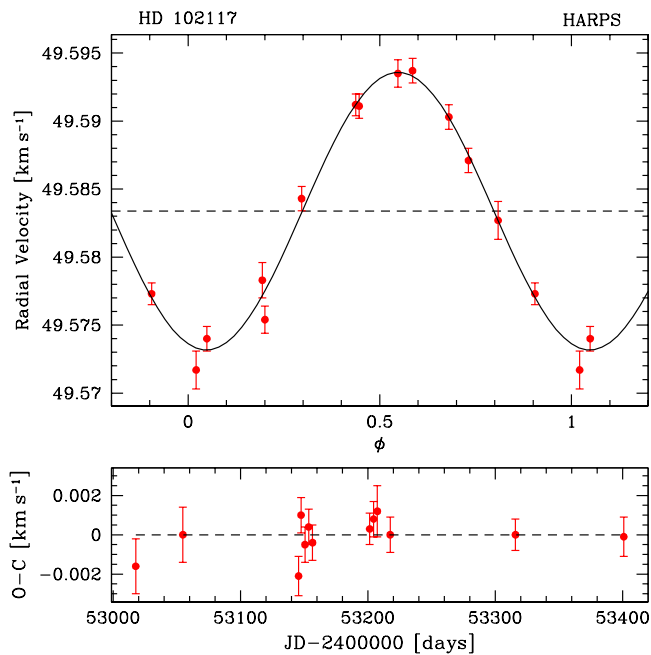}
	\caption{Phased radial velocities for HD~102117. The planet has a minimum mass of 0.14 $M_{\mathrm{Jup}}$ and an orbital period of 20.7 days.}
	\label{FigHD102117}
\end{figure}

The best-fit Keplerian orbit explaining the RV variations has a period $P$ = 20.67 days, a semi-amplitude $k$ = 10.2 m~s$^{-1}$ and an eccentricity $e$ equal to 0. We derive a companion minimum mass $m_2 \sin{i}$ = 0.14 $M_{\mathrm{Jup}}$, which is one of the lightest extrasolar planets known to date. The weighted rms around the fit is only 0.9 m~s$^{-1}$, equal to internal errors. This shows that intrinsic stellar effects, such as activity-related jitter, are negligible for this star at a level of a few tens of cm~s$^{-1}$. HD~102117~b has been previously announced by \citet{tinney04}, who find orbital parameters very similar to ours, although with slightly larger uncertainties due to the larger error bars (~$\sigma_{\mathrm{O-C}}$~=~3.3 m~s$^{-1}$~). The combination of both data sets show that the radial velocities match very well in phase and amplitude. Fig.\ref{FigHD102117} shows the phased radial velocities for HD~102117 with the corresponding residuals around the fit, and Table \ref{TablePlanets} gives the best-fit parameters for HD~102117~b.

\begin{table}
\caption{Radial-velocity values and error bars for HD~102117. All data are relative to the solar system barycenter.}
\label{TableRV_HD102117}
\centering
\begin{tabular}{c c c}
\hline\hline
\bf JD-2400000 & \bf RV & \bf Uncertainty \\
 & \bf [km~s$^{-1}$] & \bf [km~s$^{-1}$] \\
\hline
53017.85639 & 49.5717 & 0.0014 \\
53054.80495 & 49.5827 & 0.0014 \\
53145.59708 & 49.5754 & 0.0010 \\
53147.57142 & 49.5843 & 0.0009 \\
53150.67935 & 49.5911 & 0.0009 \\
53153.55336 & 49.5937 & 0.0009 \\
53156.57259 & 49.5871 & 0.0009 \\
53201.49891 & 49.5773 & 0.0008 \\
53204.47977 & 49.5740 & 0.0009 \\
53207.46689 & 49.5783 & 0.0013 \\
53217.52802 & 49.5903 & 0.0009 \\
53315.86091 & 49.5912 & 0.0008 \\
53400.83082 & 49.5935 & 0.0010 \\
\hline
\end{tabular}
\end{table}

\section{Discussion and conclusion}

Table \ref{TablePlanets} summarizes all orbital and physical properties of the planets presented in this paper. These new discoveries have been made possible thanks to the high RV precision reached with the HARPS spectrograph. This precision is demonstrated by the very low residuals around the orbital fits, allowing an accurate determination of the orbital parameters even with few data points, provided the phase coverage is good enough.

All three stars discussed in this paper are more metal-rich than the Sun and therefore confirm the observational trend showing a clear correlation between planet occurrence and metallicity.

As already mentioned in Sect. 1, these new Saturn-mass planets are located at intermediate distances from their parent star (0.1-0.5 AU). The existence of planets in this area of the mass-period diagram might pose a challenge to the standard planet formation theories. The difficulty arises from the fast runaway accretion and migration processes that would make such planets more massive and come closer to their parent stars, thereby becoming hot Jupiters (or hot Saturns). A reliable mechanism permitting termination of the accretion process at intermediate distances from the star remains to be found, although the evaporation of the disk and the metallicity of the central star might play an important role. Among the three planets presented in this paper, HD~102117~b is certainly the most challenging one in that respect, with a mass of $\sim$50 $M_{\oplus}$ and a semi-major axis of 0.15 AU. The recently discovered Neptune-mass planet around $\mu$ Ara \citep{santos04a}, orbiting at 0.09 AU, shows similar properties, but in that case the presence of other massive planets in the system might have had an influence on its final mass. On the contrary, no other massive planets seem to exist around HD~102117, at least within $\sim$5 AU. From the observational point of view, these new discoveries might indicate that a population of Neptune- and Saturn-mass planets remains to be discovered below 1 AU. The increasing precision of the radial-velocity surveys will help answer this question in the near future, thereby providing useful new constraints on planet formation theories.

\begin{acknowledgements}

We would like to thank the Swiss National Science Foundation (FNRS) and the Portuguese Funda\c c\~ao para a Ci\^encia e Tecnologia for their continuous support of this project. This study also benefitted from the support of the HPRN-CT-2002-O0308 European programme.

\end{acknowledgements}

\bibliographystyle{aa}
\bibliography{biblio}

\end{document}